\documentclass{aa}
\usepackage{graphicx,natbib}
\bibpunct{(}{)}{;}{a}{}{,}

\newcommand{\lesssim}{\mathrel{\hbox{\rlap{\hbox{\lower4pt\hbox{$\sim$}}}\hbox{$<$}}}}
\newcommand{\gtrsim}{\mathrel{\hbox{\rlap{\hbox{\lower4pt\hbox{$\sim$}}}\hbox{$>$}}}}
\newcommand{\beq}{\begin{equation}}
\newcommand{\eeq}{\end{equation}}
\newcommand{\beqa}{\begin{eqnarray}}
\newcommand{\eeqa}{\end{eqnarray}}

\def\nh{N_{\rm H}}

\def\lhx{L_{\rm hx}}
\def\lx{L_{\rm x}}
\def\vm{V_{\rm m}}
\def\lb{L_\ast}

\begin{document}

\title{Hard X-ray luminosity function and absorption distribution of
nearby AGN: INTEGRAL all-sky survey}

\author{S.~Sazonov\inst{1,2} \and M.~Revnivtsev\inst{1,2} \and
  R.~Krivonos\inst{2,1} \and E.~Churazov\inst{1,2} \and R.~Sunyaev\inst{1,2}}

\offprints{sazonov@mpa-garching.mpg.de}

\institute{Max-Planck-Institut f\"ur Astrophysik,
           Karl-Schwarzschild-Str. 1, D-85740 Garching bei M\"unchen,
           Germany
     \and   
           Space Research Institute, Russian Academy of Sciences,
           Profsoyuznaya 84/32, 117997 Moscow, Russia
}
\date{Received / Accepted}

\authorrunning{Sazonov et al.}
\titlerunning{Hard X-ray luminosity function of nearby AGN}

\abstract
{}
  {We study the hard X-ray luminosity function and absorption 
  distribution of local ($z\lesssim 0.1$) active galactic nuclei
  (AGN) and discuss the implications for AGN cosmological evolution
  and for the cosmic X-ray background (CXB). 
  }
  {We use the INTEGRAL all-sky hard X-ray survey to perform a
  statistical study of a representative sample of nearby AGN. Our
  entire all-sky sample consists of 127~AGN, of which 91 are
  confidently detected ($>5\sigma$) on the time-averaged map obtained
  with the IBIS/ISGRI instrument and 36 are detected only during
  single observations. Among the former there are 66 non-blazar AGN
  located at $|b|>5^\circ$, where the survey's identification
  completeness is $\sim 93$\%, which we use 
  for calculating the AGN luminosity function and X-ray
  absorption distribution.
  }
  {In broad agreement with previous studies, we find that the fraction
  $f_{\rm a}$ of obscured ($\log\nh>22$) objects is much higher ($\sim
  70$\%) among the low-luminosity AGN ($\lhx<10^{43.6}$~erg~s$^{-1}$)
  than among the high-luminosity ones ($\lhx>10^{43.6}$~erg~s$^{-1}$),
  $f_{\rm a}\sim 25$\%, where $\lhx$ is the luminosity in the
  17--60~keV energy band. We also find that locally the fraction of
  Compton-thick AGN is less than 20\% unless there is a significant
  population of AGN that are so strongly obscured that their observed hard
  X-ray luminosities fall below $\sim
  10^{40}$--$10^{41}$~erg~s$^{-1}$, the effective limit of our 
  survey. The constructed hard X-ray luminosity function has a
  canonical, smoothly broken power-law shape in the range
  $40<\log\lhx<45.5$  with a characteristic luminosity of
  $\log\lb=43.40\pm 0.28$. The estimated local luminosity density due
  to AGN with $\log\lhx>40$ is $(1.4\pm 0.3)\times
  10^{39}$~erg~s$^{-1}$~Mpc$^{-3}$ (17--60~keV). We demonstrate that
  the spectral shape and amplitude of the CXB are consistent with the
  simple scenario in which the $\nh$ distribution of AGN (for a given
  $\lhx/\lb(z)$ ratio) has not changed significantly since $z\sim 1.5$,
  while the AGN luminosity function has experienced pure luminosity evolution.
  }
  {}

\maketitle

\section{Introduction}
\label{s:intro}

Recent deep extragalatic X-ray surveys (see \citealt{brahas05} for a
review) have significantly advanced our understanding of the
cosmological evolution of 
active galactic nuclei (AGN) and the growth of massive black
holes. In particular, these surveys have revealed the
phenomenon of AGN ``downsizing'' , i.e. gradual transition from the
dominance of luminous quasars at high redshift ($z\gtrsim 2$) to that
of lower luminosity AGN at $z\lesssim 1$. Another important result is
the discovery of numerous type~2 (i.e. with significant X-ray
absorption) AGN at medium and high redshifts. Despite these 
successes, the census of AGN in the Universe may still be
significantly incomplete, in particular because deep X-ray surveys, performed
effectively at energies below $\sim 8$~keV, are biased against
detecting heavily obscured sources. Furthemore, since deep
extragalatic X-ray surveys usually cover small areas on the sky, they
do not probe the bright end of the AGN luminosity distribution at low redshift
($z\lesssim 0.3$). It is in these two overlapping areas -- heavily
obscured AGN and luminous nearby AGN -- where deep X-ray surveys could
be efficiently complemented by all-sky hard X-ray (above $\sim
10$~keV) surveys.

Recently, a serendipitous all-sky survey in the 3--20~keV energy band
was performed with the RXTE observatory (the RXTE Slew Survey, or
XSS), in which $\sim 100$ AGN were detected at $|b|>10^\circ$
\citep{revetal04}. This sample was used to construct an X-ray
luminosity function of local ($z\lesssim 0.1$) AGN and 
to study their absorption distribution \citep{sazrev04}. However,
despite its harder X-ray range compared to traditional X-ray surveys,
the XSS was still biased against finding sources with absorption
columns $\nh\gtrsim 10^{23}$~cm$^{-2}$. 

A comparable or better (for heavily obscured sources and/or in
crowded sky regions) capability for studying bright ($\gtrsim 1$~mCrab)
hard X-ray sources is now provided by the IBIS telescope
\citep{ubeetal03} of the INTEGRAL observatory \citep{winetal03}, owing
to its good sensitivity above 20~keV, large field of view ($\sim
28^\circ \times 28^\circ$), and good ($\sim 10^\prime$) angular
resolution. Since its launch in October 2002, INTEGRAL had observed
$\sim 60$\% of the sky by the spring of 2005, mostly around the
Galactic plane and the Galactic center. This motivated us to initiate
a series of pointed INTEGRAL observations (Proposal
0320108) aimed at covering the remaining ``empty'' (extragalactic) 
sky regions. This campaign has now been completed and practically full
coverage of the sky has been achieved after all public and
our proprietary INTEGRAL observations were added. 

Based on the all-sky hard X-ray map obtained and the follow-up source
identifications, we compiled a catalog of detected sources
\citep{krietal06}. In the present paper we use the INTEGRAL all-sky
survey to study such key statistical properties of the local AGN
population as the hard X-ray luminosity function and absorption
distribution. We also discuss the implications for the
cosmological evolution of AGN and for the cosmic X-ray background. We
note that there have been previous attempts to use INTEGRAL for
studying the statistics and demography of nearby AGN
\citep{krietal05,basetal06,becetal06}. The significant advantage of
our present study is that it is based on an all-sky, largely
serendipitious survey, which allowed us to significantly alleviate the 
previously significant bias associated with INTEGRAL AGN pointings. We
also note that a similar all-sky hard X-ray survey has since recently
been carried out by the Swift observatory \citep{maretal05}. 

\section{The AGN sample}
\label{s:sample}

\begin{table*}
\caption{INTEGRAL all-sky survey AGN sample
\label{tab:agn}
}
\smallskip

\scriptsize

\begin{tabular}{lcrrcrrcr}
\hline
\hline
\multicolumn{1}{c}{Object} &
\multicolumn{1}{c}{Class$^{\rm a}$} &
\multicolumn{1}{c}{$z$} &
\multicolumn{1}{c}{$D$} &
\multicolumn{1}{c}{Note$^{\rm b}$} &
\multicolumn{1}{c}{$F_{\rm 17-60~keV}$} & 
\multicolumn{1}{c}{$\log L_{\rm hx}$} &
\multicolumn{1}{c}{$N_{\rm H}$} & 
\multicolumn{1}{c}{Ref$^{\rm c}$} \\
   
\multicolumn{1}{c}{(sources at $|b|<5^\circ$ are given in bold)} &     
\multicolumn{1}{c}{} &     
\multicolumn{1}{c}{} &
\multicolumn{1}{c}{Mpc} & 
\multicolumn{1}{c}{} &     
\multicolumn{1}{c}{10$^{-11}$~erg/s/cm$^{2}$} &       
\multicolumn{1}{c}{erg/s} &     
\multicolumn{1}{c}{10$^{22}$cm$^{-2}$} &
\multicolumn{1}{c}{} \\
\hline
\multicolumn{9}{c}{\normalsize AGN detected at $>5\sigma$ on the time-averaged IBIS/ISGRI
map} \\
\hline
Mrk 348                                      &   S2 & 0.0150 &      &  &  $7.41\pm0.83$ & 43.52 &    30   & 2\\
IGR J01528$-$0326=MCG -01-05-047             &   S2 & 0.0172 &      & 1&  $1.63\pm0.30$ & 42.98 &         &  \\
NGC 788                                      &   S2 & 0.0136 &      &  &  $4.83\pm0.29$ & 43.25 &    44   & 3\\
LEDA 138501                                  &   S1 & 0.0492 &      &  &  $3.96\pm0.72$ & 44.30 &  $<1$   & 4\\
Mrk 1040                                     & S1.5 & 0.0167 &      &  &  $4.85\pm0.81$ & 43.43 &  $<1$   & 5\\
IGR J02343+3229=NGC 973                      &   S2 & 0.0162 &      & 1&  $3.88\pm0.63$ & 43.31 &         &  \\
NGC 1068                                     &   S2 & 0.0038 & 14.4 &  &  $1.90\pm0.30$ & 41.68 &$\gtrsim10^3$ & 6\\
{\bf 4U 0241+61}                             &   S1 & 0.0440 &      &  &  $4.75\pm0.60$ & 44.28 &  $<1$   & 7\\
NGC 1142                                     &   S2 & 0.0288 &      &  &  $4.61\pm0.35$ & 43.89 &  45     & 8\\
1H 0323+342                                  &   S1 & 0.0610 &      &  &  $2.74\pm0.47$ & 44.34 &  $<1$   & 4\\
NGC 1365                                     & S1.8 & 0.0055 & 16.9 &  &  $3.30\pm0.65$ & 42.06 & $\sim 50$ & 9\\
3C 111                                       & BLRG & 0.0485 &      &  &  $7.83\pm0.87$ & 44.59 &  $<1$   &10\\
ESO 033-G002                                 &   S2 & 0.0181 &      &  &  $1.94\pm0.27$ & 43.10 &  1      &11\\
IRAS 05078+1626                              & S1.5 & 0.0179 &      &  &  $5.93\pm0.75$ & 43.58 &  $<1$   & 4\\
Mrk 3                                        &   S2 & 0.0135 &      &  &  $6.83\pm0.30$ & 43.39 &  110    & 6\\
Mrk 6                                        & S1.5 & 0.0188 &      &  &  $3.66\pm0.30$ & 43.41 & $\sim 5$ &12\\
IGR J07563$-$4137=2MASX J07561963$-$4137420  &   S2 & 0.0210 &      &2,3& $1.24\pm0.24$ & 43.04 &  $<1$   &13\\
{\bf IGR J07597$-$3842}                      &  S1.2& 0.0400 &      & 3&  $2.91\pm0.26$ & 43.98 &  $<1$   & 8\\
ESO 209-G012                                 & S1.5 & 0.0405 &      &  &  $1.66\pm0.24$ & 43.75 &  $<1$   & 4\\
{\bf Fairall 1146}                           & S1.5 & 0.0316 &      &  &  $1.62\pm0.25$ & 43.52 &  $<1$   & 4\\
IRAS 09149$-$6206                            &   S1 & 0.0573 &      &  &  $2.06\pm0.26$ & 44.16 &  $<1$   & 4\\
Mrk 110                                      & NLS1 & 0.0353 &      &  &  $5.86\pm1.14$ & 44.17 &  $<1$   & 3\\
IGR J09446$-$2636=6dF J0944370$-$263356      &   S1 & 0.1425 &      & 4&  $3.91\pm0.74$ & 45.27 &  $<1$   & 4\\
NGC 2992                                     &   S2 & 0.0077 & 30.5 &  &  $5.08\pm0.39$ & 42.76 &   1     &14\\
MCG -5-23-16                                 &   S2 & 0.0085 &      &  &  $9.77\pm0.84$ & 43.14 &  2.3    &15\\
NGC 3081                                     &   S2 & 0.0080 & 32.5 &  &  $4.62\pm0.56$ & 42.77 &   50    &14\\
ESO 263-G013                                 &   S2 & 0.0333 &      &  &  $2.15\pm0.39$ & 43.69 &   40    &16\\
NGC 3227                                     & S1.5 & 0.0039 & 20.6 &  &  $9.05\pm0.84$ & 42.67 &  $<1$   & 3\\
NGC 3281                                     &   S2 & 0.0107 &      &  &  $3.87\pm0.64$ & 42.94 &  150    &17\\
IGR J10386$-$4947=2MASX J10384520$-$4946531  &   S1 & 0.0600 &      & 5&  $1.47\pm0.24$ & 44.05 &   1     & 8\\
IGR J10404$-$4625=LEDA 93974                 &   S2 & 0.0239 &      & 6&  $2.11\pm0.35$ & 43.38 &  2.8    & 8\\
NGC 3783                                     &   S1 & 0.0097 & 38.5 &  & $12.28\pm1.86$ & 43.34 &  $<1$   & 3\\
IGR J12026$-$5349=WKK 0560                   &   S2 & 0.0280 &      &2,3&  $2.44\pm0.30$ & 43.59 &   2     &13\\
NGC 4151                                     & S1.5 & 0.0033 & 20.3 &  & $47.38\pm0.43$ & 43.37 &   8     &15\\
Mrk 50                                       &   S1 & 0.0234 &      &  &  $1.31\pm0.23$ & 43.16 &  $<1$   & 4\\
NGC 4388                                     &   S2 & 0.0084 & 16.8 &  & $17.89\pm0.31$ & 42.79 &   40    &14\\
NGC 4395                                     & S1.8 & 0.0011 &  3.6 &  &  $1.55\pm0.29$ & 40.38 &   2     &18\\
NGC 4507                                     &   S2 & 0.0118 &      &  & $10.93\pm0.49$ & 43.48 &   59    &15\\
NGC 4593                                     &   S1 & 0.0090 & 39.5 &  &  $5.86\pm0.25$ & 43.04 &  $<1$   & 3\\
NGC 4945                                     &   S2 & 0.0019 &  5.2 &  & $19.92\pm0.36$ & 41.81 & 220     & 6\\
ESO 323-G077                                 & S1.2 & 0.0150 &      &  &  $2.78\pm0.35$ & 43.09 &  30     & 8\\
IGR J13091+1137=NGC 4992                     &      & 0.0251 &      & 2&  $3.49\pm0.41$ & 43.65 &  90     &13\\
IGR J13149+4422=Mrk 248                      &   S2 & 0.0366 &      & 7&  $2.16\pm0.38$ & 43.77 &         &  \\
Cen A                                        & NLRG & 0.0018 &  4.9 &  & $56.09\pm0.32$ & 42.21 &   11    &15\\
MCG -6-30-15                                 & S1.2 & 0.0077 &      &  &  $3.62\pm0.38$ & 42.62 &  $<1$   & 5\\
Mrk 268                                      &   S2 & 0.0399 &      &  &  $1.74\pm0.30$ & 43.76 &         &  \\
{\bf 4U 1344$-$60}                           &   S1 & 0.0130 &      & 6&  $5.96\pm0.25$ & 43.30 &  $<1$   & 3\\
IC 4329A                                     & S1.2 & 0.0160 &      &  & $16.15\pm0.51$ & 43.92 &  $<1$   & 5\\
{\bf Circinus galaxy}                        &   S2 & 0.0014 &  4.2 &  & $18.76\pm0.26$ & 41.60 & 400     & 6\\
NGC 5506                                     & S1.9 & 0.0062 & 28.7 &  & $13.34\pm0.66$ & 43.12 &   2.6   &15\\
{\bf IGR J14493$-$5534=2MASX J14491283$-$5536194}&     &        &   & 8&  $1.64\pm0.23$ &       &   10    & 8\\
{\bf IGR J14515$-$5542=WKK 4374}             &   S2 & 0.0180 &      & 3&  $1.55\pm0.22$ & 43.00 &  $<1$   & 8\\
IGR J14552$-$5133=WKK 4438                   &  NLS1& 0.0160 &      &3,9& $1.40\pm0.23$ & 42.85 &  $<1$   & 4\\
IC 4518                                      &   S2 & 0.0157 &      &  &  $2.44\pm0.23$ & 43.08 &         &  \\
WKK 6092                                     &   S1 & 0.0156 &      &  &  $1.66\pm0.24$ & 42.91 &  $<1$   &19\\
IGR J16185$-$5928=WKK 6471                   &  NLS1& 0.0350 &      &3,9& $1.73\pm0.24$ & 43.64 &         &  \\
ESO 137-G34                                  &   S2 & 0.0092 &      &  &  $1.68\pm0.23$ & 42.45 &         &  \\
IGR J16482$-$3036=2MASX J16481523$-$3035037  &   S1 & 0.0313 &      & 6&  $2.61\pm0.24$ & 43.72 &  $<1$   & 8\\
NGC 6221                                     &   S2 & 0.0050 & 19.4 &  &  $1.90\pm0.28$ & 41.94 &   1     &20\\
IGR J16558$-$5203                            & S1.2 & 0.0540 &      & 3&  $2.93\pm0.21$ & 44.26 &  $<1$   & 8\\  
NGC 6300                                     &   S2 & 0.0037 & 14.3 &  &  $4.71\pm0.45$ & 42.07 &  25     &15\\
{\bf IGR J17204$-$3554}                      &      &        &      &10&  $1.13\pm0.17$ &       &  13     &21\\
{\bf GRS 1734$-$292}                         &   S1 & 0.0214 &      &  &  $7.42\pm0.14$ & 43.83 &  $<1$   &22\\
IGR J17418$-$1212=2E 1739.1$-$1210           &   S1 & 0.0372 &      &11&  $2.55\pm0.29$ & 43.86 &  $<1$   &23\\
{\bf IGR J17488$-$3253}                      &   S1 & 0.0200 &      &3 &  $1.91\pm0.14$ & 43.19 &  $<1$   & 8\\ 
{\bf IGR J17513$-$2011}                      & S1.9 & 0.0470 &      &3 &  $2.32\pm0.17$ & 44.03 &         &  \\
{\bf IGR J18027$-$1455}                      &   S1 & 0.0350 &      &12&  $2.93\pm0.22$ & 43.87 &  $<1$   &24\\
3C 390.3                                     &  BLRG& 0.0561 &      &  &  $6.16\pm0.64$ & 44.61 &  $<1$   & 3\\
IGR J18559+1535=2E 1853.7+1534               &   S1 & 0.0838 &      &13,14&  $2.27\pm0.23$ & 44.54 &$<1$  &23\\
1H 1934$-$063                                &   S1 & 0.0106 &      &  &  $1.77\pm0.31$ & 42.59 &  $<1$   & 4\\
NGC 6814                                     & S1.5 & 0.0052 & 22.8 &  &  $4.73\pm0.41$ & 42.47 &  $<1$   & 5\\
Cygnus A                                     &  NLRG& 0.0561 &      &  &  $5.77\pm0.33$ & 44.58 &  20     &26\\
{\bf IGR J2018+4043=2MASX J20183871+4041003} &      &        &      &15&  $1.89\pm0.28$ &       &   7     & 8\\
IGR J20286+2544=MCG +04-48-002               &   S2 & 0.0142 &      &16&  $3.31\pm0.58$ & 43.12 &  50     & 8\\
Mrk 509                                      & S1.2 & 0.0344 &      &  &  $5.51\pm0.83$ & 44.13 &  $<1$   &15\\
{\bf IGR J21178+5139=2MASX J21175311+5139034}&      &        &      &17&  $1.93\pm0.34$ &       &         &  \\
{\bf IGR J21247+5058}                        &   S1 & 0.0200 &      &12&  $8.56\pm0.34$ & 43.84 &  $<1$   &24\\
{\bf IGR J21277+5656}                        &   S1 & 0.0144 &      &13&  $2.68\pm0.45$ & 43.04 &  $<1$   & 8\\
NGC 7172                                     &   S2 & 0.0087 & 33.9 &  &  $5.99\pm0.50$ & 42.92 &  13     &15\\
MR 2251$-$178                                &   S1 & 0.0640 &      &  &  $4.75\pm0.48$ & 44.62 &  $<1$   &27\\
NGC 7469                                     & S1.2 & 0.0163 &      &  &  $4.74\pm0.78$ & 43.40 &  $<1$   & 3\\
Mrk 926                                      & S1.5 & 0.0469 &      &  &  $3.56\pm0.53$ & 44.21 &  $<1$   & 3\\
\end{tabular}
\end{table*}

\setcounter{table}{0}
\begin{table*}

\caption{--continued
}
\smallskip

\scriptsize

\begin{tabular}{lcrrcrrcr}
\hline
\hline
\multicolumn{1}{c}{Object} &
\multicolumn{1}{c}{Class$^{\rm a}$} &
\multicolumn{1}{c}{$z$} &
\multicolumn{1}{c}{$D$} &
\multicolumn{1}{c}{Note$^{\rm b}$} &
\multicolumn{1}{c}{$F_{\rm 17-60~keV}$} & 
\multicolumn{1}{c}{$\log L_{\rm hx}$} &
\multicolumn{1}{c}{$N_{\rm H}$} & 
\multicolumn{1}{c}{Ref$^{\rm c}$} \\
   
\multicolumn{1}{c}{(sources at $|b|<5^\circ$ are given in bold)} &     
\multicolumn{1}{c}{} &     
\multicolumn{1}{c}{} &
\multicolumn{1}{c}{Mpc} & 
\multicolumn{1}{c}{} &     
\multicolumn{1}{c}{10$^{-11}$~erg/s/cm$^{2}$} &       
\multicolumn{1}{c}{erg/s} &     
\multicolumn{1}{c}{10$^{22}$cm$^{-2}$} &
\multicolumn{1}{c}{} \\
\hline
\multicolumn{9}{c}{\normalsize Blazars} \\
{\bf 87GB 003300.9+593328}                   &   Bl & 0.0860 &      &  &  $1.03\pm0.15$ & 44.22 &  $<1$   & 1\\
S5 0836+71                                   &   Bl & 2.1720 &      &  &  $3.47\pm0.33$ & 48.03 &  $<1$   & 1\\
PKS 1219+04                                  &   Bl & 0.9650 &      &  &  $1.30\pm0.23$ & 46.74 &  $<1$   & 4\\
3C 273                                       &   Bl & 0.1583 &      &  & $13.83\pm0.23$ & 45.92 &  $<1$   & 1\\
3C 279                                       &   Bl & 0.5362 &      &  &  $1.50\pm0.28$ & 46.18 &  $<1$   & 1\\
Mrk 501                                      &   Bl & 0.0337 &      &  &  $4.53\pm0.43$ & 44.02 &  $<1$   & 1\\
PKS 1830$-$211                               &   Bl & 2.5070 &      &  &  $3.35\pm0.22$ & 48.17 &  $<1$   &25\\
BL Lac                                       &   Bl & 0.0686 &      &  &  $1.88\pm0.38$ & 44.28 &  $<1$   & 1\\
3C 454.3                                     &   Bl & 0.8590 &      &  & $13.92\pm0.55$ & 47.64 &  $<1$   & 1\\
\hline
\multicolumn{9}{c}{\normalsize Transiently detected AGN} \\
\hline
NGC 0526A                       &  S1.5 &  0.0191 &       &  & $3.72\pm0.97$ &  43.43 & & \\
ESO 297-G018                    &  S2   &  0.0252 &       &  & $4.41\pm0.94$ &  43.75 & & \\
NGC 1052                        &  S2   &  0.0050 &  17.8 &  & $2.12\pm0.43$ &  41.91 & & \\
IGR J03334+3718                 &  S1   &  0.0547 &       &18& $1.96\pm0.41$ &  44.09 & & \\
{\bf LEDA 168563}               &  S1   &  0.0290 &       &  & $4.42\pm1.11$ &  43.88 & & \\
AKN 120                         &  S1   &  0.0327 &       &  & $8.38\pm3.52$ &  44.26 & & \\
MCG 8-11-11                     &  S1.5 &  0.0205 &       &  & $5.61\pm1.51$ &  43.67 & & \\ 
{\bf IRAS 05589+2828}           &  S1   &  0.0330 &       &  & $3.03\pm0.62$ &  43.83 & & \\ 
LEDA 096373                     &  S2   &  0.0294 &       &  & $2.71\pm0.61$ &  43.68 & & \\ 
PG 0804+761                     &  S1   &  0.1000 &       &  & $1.21\pm0.33$ &  44.43 & & \\ 
NGC 4051                        &  S1.5 &  0.0023 &  17.0 &  & $2.11\pm0.54$ &  41.87 & & \\
NGC 4138                        &  S1.9 &  0.0030 &  17.0 &  & $2.39\pm0.49$ &  41.92 & & \\
WAS 49B                         &  S2   &  0.0630 &       &  & $0.95\pm0.35$ &  43.91 & & \\ 
NGC 4253                        &  S1.5 &  0.0129 &       &  & $1.53\pm0.31$ &  42.70 & & \\
NGC 4258                        &  S1.9 &  0.0015 &   6.8 &  & $1.90\pm0.56$ &  41.03 & & \\ 
XSS J12389$-$1614=IGR J12391$-$1612 &  S2   &  0.0367 &   &19& $3.01\pm0.62$ &  43.92 & & \\ 
{\bf WKK 1263}                  &  S2   &  0.0244 &       &  & $1.28\pm0.31$ &  43.19 & & \\ 
PKS 1241-399                    &  QSO  &  0.1910 &       &  & $1.77\pm0.40$ &  45.21 & & \\ 
Mrk 783                         &  S1.5 &  0.0672 &       &  & $1.23\pm0.32$ &  44.08 & & \\ 
IGR J13038+5348=MCG 09-21-096   &  S1   &  0.0299 &       &18& $2.11\pm0.45$ &  43.58 & & \\        
NGC 5033                        &  S1.9 &  0.0029 &  18.7 &  & $1.21\pm0.28$ &  41.71 & & \\ 
ESO 383-G018                    &  S2   &  0.0124 &       &  & $1.74\pm0.38$ &  42.72 & & \\ 
NGC 5252                        &  S1.9 &  0.0230 &       &  & $2.60\pm0.78$ &  43.44 & & \\ 
IGR J14175$-$4641               &  S2   &  0.0760 &       &3 & $1.30\pm0.28$ &  44.21 & & \\
NGC 5548                        &  S1.5 &  0.0172 &       &  & $2.06\pm0.50$ &  43.08 & & \\ 
ESO 511-G030                    &  S1   &  0.0224 &       &  & $3.33\pm0.91$ &  43.53 & & \\ 
IGR J14471$-$6319               &  S2   &  0.0380 &       &3 & $0.95\pm0.23$ &  43.45 & & \\
NGC 6240                        &  S2   &  0.0245 &       &  & $4.66\pm1.38$ &  43.75 & & \\
3C 382                          &  S1   &  0.0579 &       &  & $3.96\pm1.75$ &  44.45 & & \\ 
ESO 103-G035                    &  S2   &  0.0133 &       &  & $6.64\pm1.34$ &  43.37 & & \\
XSS J19459+4508=IGR J19473+4452 &  S2   &  0.0532 &       &20& $1.52\pm0.45$ &  43.96 & & \\
3C 403                          &  NLRG &  0.0590 &       &  & $0.96\pm0.33$ &  43.85 & & \\
4C +74.26                       &  BLRG &  0.1040 &       &  & $3.93\pm1.22$ &  44.98 & & \\ 
S5 2116+81                      &  S1   &  0.0840 &       &  & $2.91\pm0.98$ &  44.65 & & \\ 
NGC 7314                        &  S1.9 &  0.0048 &       &  & $2.17\pm0.58$ &  41.99 & & \\ 
Mrk 915                         &  S1   &  0.0241 &       &  & $2.62\pm0.56$ &  43.49 & & \\       
\hline
\end{tabular}
\smallskip


$^{\rm a}$ Optical AGN class: Bl -- blasar (BL Lac object or flat-spectrum
radio quasar), S1, S1.2, S1.5, S1.8, S1.9, S2 -- Seyfert galaxy, NLS1 -- narrow-line Seyfert 1
galaxy, BLRG -- broad-line radio galaxy, NLRG --  narrow-line radio
galaxy, QSO -- quasar.

$^{\rm b}$ Note or reference for INTEGRAL discovered AGN: (1) \citet{buretal06b}, (2)
\citet{sazetal05}, (3) \citet{masetal06d}, (4) galaxy=1RXS 
J094436.5$-$263353, a Seyfert~1 nucleus is suggested by the 6dF
spectrum, (5) \citet{moretal06}, =SWIFT J1038.8$-$4942  
(6) \citet{masetal06}, (7) SDSS spectrum indicates a Seyfert~2 nucleus,
(8) Swift localization consistent with this galaxy, (9)
\citet{revetal06}, (10) \citet{basetal05}, (11) \citet{toretal04},
(12) \citet{masetal04}, (13) \citet{biketal06}, (14)
\citet{masetal06b} (15) \citet{kenetal06}, (16) 
\citet{masetal06c}, (17) \citet{basetal06}, (18) \citet{buretal06}, (19) =2MASX
J12390630$-$1610472 \citep{sazetal05,masetal06}, (20) =2MASX
J19471938+4449425 \citep{sazetal05,biketal06,masetal06b}.

$^{\rm c}$ Quoted $\nh$ value or limit is adopted from or based on: (1)
\cite{donetal05}, (2) RXTE data, (3) ASCA data, (4) ROSAT data, (5) \citet{reynolds97},
(6) \citet{matetal00}, (7) \citet{maletal97}, (8) Swift data, (9)
\citet{risetal05}, $N_{\rm H}$ is strongly variable, (10)
\citet{lewetal05}, (11) \citet{vigetal98}, (12) \citet{immetal03},
complex X-ray absorption, (13) \citet{sazetal05}, (14)
\citet{risetal02}, (15) \citet{sazrev04}, (16) Swift data, it is also
possible that the X-ray spectrum is reflection-dominated with $N_{\rm
H}>10^{24}$ \citep{ajeetal06}, (17)
\citet{vigetal02}, (18) \citet{moretal05}, (19) \citet{wouetal98},
(20) \citet{levetal01}, (21) \citet{basetal05}, (22)
\citet{sazetal04b}, (23) Einstein data, (24) assumed upper limit based on
Seyfert~1 classification, (25) \citet{oshetal01}, (26) \citet{youetal02},
(27) \citet{reetur00}. 

\end{table*}

The catalog of \citet{krietal06} includes in total 127 AGN, which are
listed in Table~\ref{tab:agn}. Of these, 91 AGN (referred to below  as
the main sample) were detected with more than $5\sigma$ significance
on the time-averaged all-sky map obtained with IBIS/ISGRI in the
17--60~keV energy band and are thus well-suited for statistical
analysis. In addition, there are 36 AGN, listed in the lower part of
Table~\ref{tab:agn}, that were confidently detected in single INTEGRAL
orbits or observations ($\sim$1--3~days long) but fall below
the 5$\sigma$ threshold on the average map. We exclude these
transiently detected AGN from the following statistical analysis. 

Most of the AGN presented were known as such before INTEGRAL, while 33 sources
(those with ``IGR'' names) have been identified as AGN primarily based
on their hard X-ray detection with INTEGRAL. Two of these newly
discovered AGN -- IGR~J09446$-$2636 and IGR~J13149+4422 -- are
reported here for the first time. As indicated in column~2 of
Table~\ref{tab:agn}, 77 objects in our main sample are optically
classified as Seyfert galaxies (4 of them radio loud) and 9 are
blazars (BL Lac objects or flat-spectrum radio quasars). Although the
optical class has not yet been reported for 5 objects, their X-ray,
optical, infrared, and radio properties strongly suggest that they are
all Seyfert galaxies. The quoted optical classes are based on the NED unless a
reference is provided. Of the AGN at $|b|>10^\circ$ in our main
sample, 30 were also detected during the RXTE Slew Survey.

\begin{figure}
\centering
\includegraphics[width=\columnwidth]{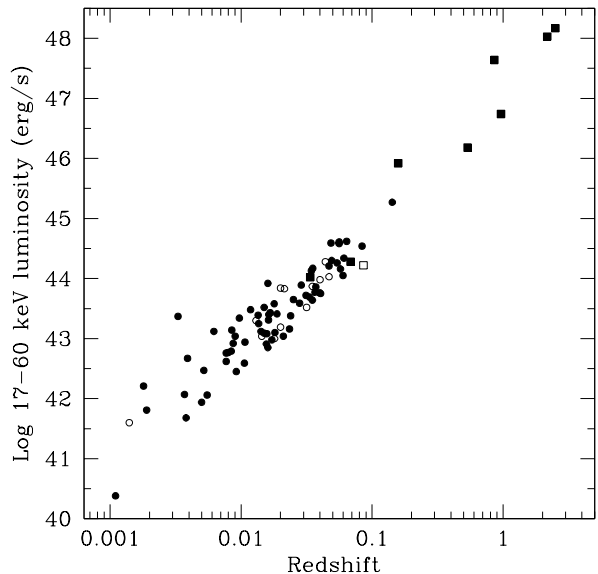}
\caption{Hard X-ray luminosity vs. redshift for the identified AGN
from the INTEGRAL all-sky survey (the main sample). Circles and
squares indicate emission-line AGN and blazars, respectively. Empty
symbols indicate AGN located at $|b|<5^\circ$, which are not used in
the statistical analysis. 
}
\label{fig:z_lum}
\end{figure}

We found published redshifts for all but 4 of our AGN (column~3 of
Table~\ref{tab:agn}) and used these to determine the source
luminosities in the observed 17--60~keV energy band (column~7) from the
hard X-ray fluxes measured by IBIS/ISGRI (column~6). For the nearest
sources ($z\lesssim 0.01$) the distances from the Nearby 
Galaxies Catalogue \citep{tully88} were adopted (column~4), while
the luminosity distances to the other objects were determined from their
redshifts assuming a cosmology with $\Omega_{\rm m}=0.3$,
$\Omega_\Lambda=0.7$, and $H_0=75$~km~s$^{-1}$~Mpc$^{-1}$, which is
used throughout the paper. Figure~\ref{fig:z_lum} shows the 
distribution of our AGN (of the main sample) in the
redshift-luminosity plane. One can see that all of the emission-line
(i.e. non-blazar) AGN are located at 
$z<0.1$, except for IGR~J09446$-$2636 at $z\approx 0.14$, while there
are several blazars as distant as $z\sim$1--2.5, with (isotropical)
luminosities up to $\sim 10^{48}$~erg~s$^{-1}$. Because of the
proximity of the non-blazar AGN, our quoted values for their observed
luminosities may also be regarded as their rest-frame 17--60~keV
luminosities.

We also collected information on the X-ray absorption columns
for the AGN of the main sample (column~8 of Table~\ref{tab:agn}). For
this purpose we 
either used the literature or analyzed publicly available spectral
data from different X-ray instruments, including ASCA/GIS,
Einstein/IPC, RXTE/PCA, and Swift/XRT. The spectral modelling was done
with XSPEC v12 \citep{arnaud96}. The X-ray spectra (usually above
2~keV) were fitted by a model consisting of a power law photoabsorbed
by a column $\nh$ of neutral material and a fluorescent neutral-iron
emission line at rest energy 6.4 keV if the latter was required by the fit
(model ${\rm zphabs\times(powerlaw+Gaussian)}$ in XSPEC). If the quality of the
data allowed, we considered the power-law slope a free parameter;
otherwise, the photon index was fixed at the canonical
(e.g. \citealt{reynolds97,turetal97}) AGN value $\Gamma=1.8$. Note
that we did not try to evaluate $\nh$ if it was clear that
$\nh<10^{22}$~cm$^{-2}$ (after subtraction of the Galactic absorption),
as we regard such sources as unobscured throughout the paper. Finally,
for several Seyfert~1 galaxies in our sample, which are expected to
have little X-ray absorption, we used ROSAT/PSPC soft X-ray data to
verify that indeed $\nh<10^{22}$~cm$^{-2}$ for them.

\subsection{Unidentified sources}
\label{s:unident}
 
There are 24 unidentified sources among those detected on the average
IBIS/ISGRI map \citep{krietal06}, so our all-sky AGN sample can be
significantly incomplete. However, since most of these objects are located
close to the Galactic plane, we can optimize our statistical
analysis by excluding the Galactic plane region 
$|b|<5^\circ$ from consideration. This leads to only a small 
decrease in the number of identified AGN -- to 74, but to a dramatic 
reduction of the number of unidentified sources -- to 7. The
coordinates and fluxes of these high Galactic-latitude sources -- AGN
candidates -- are given in Table~\ref{tab:noid}. We note that the
apparently high sky density of unidentified sources in the direction
$l\sim 330^\circ$, $b\sim 10^\circ$ is probably the result of the
extensive INTEGRAL observations of the Galactic plane and of such
interesting sources as Cen~A, NGC 4945, and SN~1006.

The treatment that follows is thus based on the $|b|>5^\circ$ AGN
sample. Furthermore, we restrict our consideration to the
emission-line AGN (66 in total) and so do not discuss blazars anymore.

\begin{table*}
\caption{Unidentified INTEGRAL sources at $|b|>5^\circ$
\label{tab:noid}
}
\smallskip

\scriptsize

\begin{tabular}{lrrrrrl}
\hline
\hline
\multicolumn{1}{c}{Source} &
\multicolumn{1}{c}{RA} &
\multicolumn{1}{c}{Dec.} &
\multicolumn{1}{c}{l} &
\multicolumn{1}{c}{b} &
\multicolumn{1}{c}{$F_{\rm 17-60~keV}$} & 
\multicolumn{1}{c}{Likely associations}\\

\multicolumn{1}{c}{} &
\multicolumn{2}{c}{(J2000.0)} &
\multicolumn{2}{c}{} &
\multicolumn{1}{c}{10$^{-11}$~erg/s/cm$^{2}$} &
\multicolumn{1}{c}{} \\
\hline
IGR J02466-4222  &  41.65 & -42.37 & 253.48 & -62.07 & $3.08\pm0.55$& \\
IGR J09522-6231  & 148.05 & -62.52 & 283.83 &  -6.50 & $1.17\pm0.21$& \\
IGR J13107-5551  & 197.68 & -55.87 & 305.66 &   6.90 & $1.85\pm0.30$& XMMSL1 J131042.6-555206 \citep{reaetal05}=PMN J1310-5552 \\
IGR J14561-3738  & 224.04 & -37.65 & 328.61 &  18.93 & $1.40\pm0.26$& XSS J14562-37 \\
IGR J16500-3307  & 252.51 & -33.10 & 349.73 &   7.34 & $1.61\pm0.23$& 1RXS J164955.1-330713 \\
IGR J16562-3301  & 254.06 & -33.03 & 350.60 &   6.37 & $1.97\pm0.21$& SWIFT J1656.3-3302 \citep{tueetal06} \\
IGR J17350-2045  & 263.75 & -20.76 &   5.69 &   6.36 & $1.29\pm0.18$& \\
\hline
\end{tabular}

\end{table*}

\section{X-ray absorption distribution}
\label{s:nhdist}

We first address the X-ray absorption distribution of nearby
($z\lesssim 0.1$) emission-line AGN. One of the most important
findings of the RXTE Slew Survey was that the fraction of
significantly X-ray absorbed AGN ($\log\nh>22$) drops with increasing
luminosity \citep{sazrev04}. Motivated by this result, we divided our
current INTEGRAL sample into two parts: AGN with $\log\lhx<43.6$ and
those with $\log\lhx>43.6$, where $\lhx$ is the luminosity in the 17--60~keV
band. The chosen dividing luminosity approximately 
corresponds to the value of the 3--20~keV luminosity ($\log\lx=43.5$)
that we used in the RXTE case given that typically 
$\log(\lhx/\lx)\sim 0.1$ for local AGN, as will be shown in \S\ref{s:3-20} 
below. This boundary also approximately corresponds to the bend of
the hard X-ray luminosity function (\S\ref{s:lumfunc}). The resulting
low- and high-luminosity samples include 42 and 24 objects,
respectively. 

\begin{figure}
\centering
\includegraphics[width=\columnwidth]{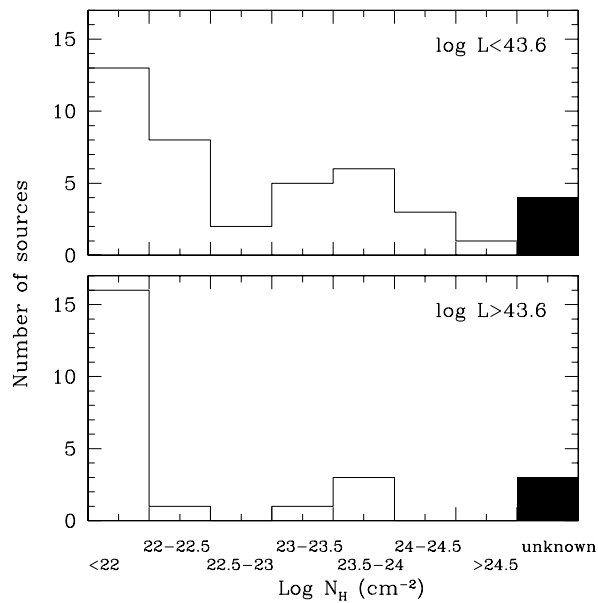}
\caption{Observed X-ray absorption distribution of the low-luminosity AGN
(top panel), and high-luminosity AGN (bottom panel). The shaded part
of each diagram shows the number of AGN with unknown $\nh$.
}
\label{fig:nhdist_lum}
\end{figure}

Figure~\ref{fig:nhdist_lum} shows the observed $\nh$ distributions for
these subsamples. It can be seen that, while $\sim 66$\% of
local $\log\lhx<43.6$ AGN are obscured ($\log\nh>22$), the
corresponding fraction is only $\sim 24$\% among the higher
luminosity AGN. This result confirms our XSS finding but it is
obtained here in a more straightforward manner. In the INTEGRAL case,
the efficiency of source detection at energies above $\sim 
20$~keV is practically independent of intrinsic
source absorption, as long as $\log\nh\lesssim 24.5$. In the XSS case,
the detection was based on 3--20~keV fluxes, so we had to correct the observed
$\nh$ distribution for the significant loss in sensitivity to sources
with $\log\nh>23$.

We note that the above conclusion with respect to the fraction of
obscured AGN as a function of luminosity is unlikely to change
once the absorption columns are measured for those INTEGRAL AGN for
which this information is not available yet, since there are only 4
and 3 such sources out of the 42 low- and 24 high-luminosity AGN,
respectively (see Fig.~\ref{fig:nhdist_lum}). 
 
We point out that the observed (see Fig.~\ref{fig:nhdist_lum}) small
fraction of substantially Compton-thick AGN -- sources with
$\log\nh\gtrsim 24.5$ -- should be interpreted with caution, since in 
their case the intrinsic hard X-ray flux is expected to be
substantially suppressed by Comptonization on the cold material obscuring the
AGN. The discussion of the fraction of Compton-thick AGN in the local
Universe will be continued in Sect.~\ref{s:summary}.

\section{Hard X-ray luminosity function}
\label{s:lumfunc}

\begin{figure}
\centering
\includegraphics[width=\columnwidth]{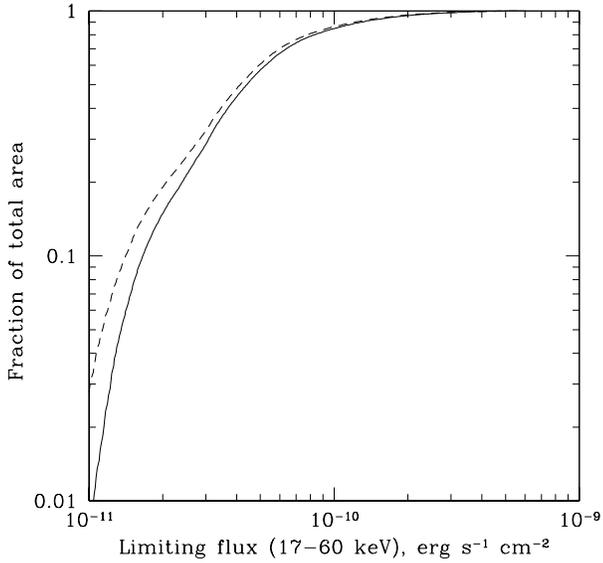}
\caption{Fraction of the total sky area as a function of survey
sensitivity (the dashed line). The solid line shows the corresponding
fraction for the $|b|>5^\circ$ sky. 
}
\label{fig:area_flux}
\end{figure}

We now define the hard X-ray luminosity function of nearby
emission-line AGN as $\phi(\lhx)\equiv dN_{\rm AGN}/d\log\lhx$.
We first estimated $\phi(\lhx)$ in binned form using the $1/\vm$ method
\citep{schmidt68}, i.e. by summing $1/\vm (L_{\rm hx,i})$ values for
our AGN in specified luminosity intervals, 
where $\vm(L_{\rm hx,i})$ is the space volume in which an AGN with
luminosity $L_{\rm hx,i}$ could be detected by the survey. The $\vm$ 
calculation was based on the dependence of the sky area covered by
the survey on the achieved sensitivity, which is shown in
Fig.~\ref{fig:area_flux}. We note that $\sim 75$\% and $\sim 50$\% of the
$|b|>5^\circ$ sky have been covered down to 5 and 3~mCrab,
respectively, where 1~mCrab corresponds to $\sim 1.4\times
10^{-11}$~erg~s~cm$^{-2}$ (17--60~keV). The corresponding fractions
are somewhat higher for the whole sky due to the overexposure of the
Galactic plane region by INTEGRAL.

\begin{figure}
\centering
\includegraphics[width=\columnwidth]{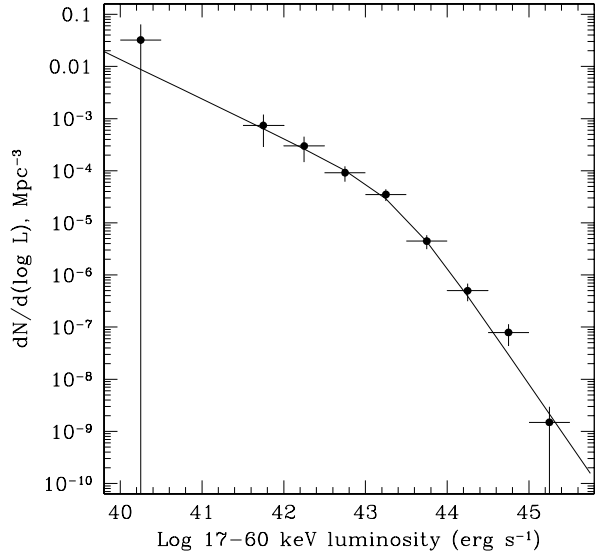}
\caption{Hard X-ray luminosity function of local emission-line AGN
obtained with INTEGRAL in binned form (solid circles and 1$\sigma$
error bars). The solid line shows the analytic approximation given by
Eq.~(\ref{eq:lum_model}) and Table~\ref{tab:lumfunc}. 
}
\label{fig:lumfunc}
\end{figure}

We performed the calculation in the luminosity range $40<\log\lhx<45.5$,
which includes all of our emission-line AGN, 66 objects in total. The
resulting binned luminosity function with estimated 1$\sigma$
statistical errors is shown in Fig.~\ref{fig:lumfunc}.

We next sought an analytic representation of the luminosity function
in the form of a smoothly broken power-law model, as is conventional
in AGN research:
\beq
\phi(\lhx)=\frac{A}{(\lhx/\lb)^{\gamma_1}+(\lhx/\lb)^{\gamma_2}}.
\label{eq:lum_model}
\eeq
For this purpose we used the maximum likelihood estimator 
\beq
L=-2\sum_{i}\ln\frac{\phi(L_{\rm hx,i})\vm(L_{\rm hx,i})}
{\int\phi(\lhx)\vm(\lhx)\,d\log\lhx},
\label{eq:like_lum}
\eeq
where $i$ runs over all sampled AGN.
 
By minimizing $L$, we found the best-fit values for the break
luminosity $\lb$, as well as for the slopes $\gamma_1$ and
$\gamma_2$. Since the maximum likelihood method directly determines
only the shape of the function, we derived the normalization $A$
by requiring that the number of AGN predicted by the model be equal to
the number of AGN in the sample. The resulting best-fit model is shown in
Fig.~\ref{fig:lumfunc} and its parameters are given in
Table~\ref{tab:lumfunc} together with their estimated 1$\sigma$ errors; the
error for the coefficient $A$ is not quoted, since this parameter
is strongly correlated with the others. The analytic fit is apparently
in good agreement with the binned $\phi(\lhx)$, and the good quality of
the fit is confirmed by the Kolmogorov--Smirnov test. 
 
\begin{table}
\caption{Parameters of the hard X-ray luminosity function
\label{tab:lumfunc}
}
\smallskip

\begin{tabular}{l|l}
\hline
\hline
Parameter & Value and 1$\sigma$ range\\
\hline
$\log L_\ast$ & 43.40 ($43.12\div 43.68$) \\
$\gamma_1$ & 0.76 ($0.56\div 0.94$) \\
$\gamma_2$ & 2.28 ($2.06\div 2.56$) \\
$A$ (Mpc$^{-3}$) & $3.55\times 10^{-5}$ \\ 
$n_{\rm 17-60~keV}(>40)$ ($10^{-3}$ Mpc$^{-3}$) & 9 ($4\div 18$) \\
$n_{\rm 17-60~keV}(>41)$ ($10^{-3}$ Mpc$^{-3}$) & 1.4 ($0.9\div 2.0$) \\   
$\rho_{\rm 17-60~keV}(>40)$ ($10^{38}$~erg~s$^{-1}$~Mpc$^{-3}$)
 & 14.1 ($11.8\div 17.1$) \\
$\rho_{\rm 17-60~keV}(>41)$ ($10^{38}$~erg~s$^{-1}$~Mpc$^{-3}$)
 & 12.4 ($11.0\div
14.0$) \\
\hline
$P_{\rm KS}$ & $>0.9$\\
\hline
\end{tabular}

\end{table}

\subsection{Cumulative number density and luminosity density}
\label{s:emis}

By integrating the derived luminosity function over luminosity, we can
find the total number density of nearby emission-line AGN with
$\log\lhx>40$:
\beq
n(>40)=\int_{40}^\infty\phi(\lhx)\,d\log\lhx.
\eeq
The resulting value is given in Table~\ref{tab:lumfunc}, together with
its 1$\sigma$ uncertainty range. It should be noted that this
cumulative number density is strongly dominated by AGN at the low
end of the probed luminosity range and that there is only one AGN in
our sample (NGC~4395) with $\log\lhx<41$. For this reason we also provide
our estimate of (similarly defined) $n(>41)$ in Table~\ref{tab:lumfunc}.

We can also determine the cumulative luminosity density of nearby
emission-line AGN with $\log\lhx>40$:
\beq
\rho_{\rm 17-60~keV}(>40)=\int_{40}^\infty\phi(\lhx)\lx\,d\log\lhx. 
\eeq
The result is presented in Table~\ref{tab:lumfunc}. Also given in the
table is the value of $\rho_{\rm 17-60~keV}(>41)$, which is better
constrained by the data. 
 
\subsection{Systematic uncertainties}

There are some systematic effects that may affect the luminosity
function obtained. First, our input AGN sample may be incomplete
because there are 7 unidentified INTEGRAL sources at $|b|>5^\circ$,
some of which may eventually prove to be emisison-line
AGN. Assuming that there is no strong luminosity bias for these
sources, the amplitude of our luminosity function can be 
underestimated due to incompleteness by $\sim 10$\% at most.

Another cause of concern is that 15 of our $|b|>5^\circ$
emission-line AGN have been at least once the target of a pointed
INTEGRAL observation, i.e. we are not dealing with a truly
serendipitous survey. In order to assess the maximum possible systematic
effect associated with these AGN pointings, the following argument
can be suggested. Suppose that an area $S$ of the sky was
covered by the survey down to flux $f_1$ or lower, 
while the total area of the survey is $S_0$. Suppose next that a total
of $N$ sources with flux $f_1<f<f_2$ were detected, of which $N_{\rm t}$
were observational targets. Then the sky density $\rho$ of sources
with $f_1<f<f_2$ can be expected to be bound in the range ($\rho_{\rm
min} <\rho<\rho_{\rm max})$:
\beq
\frac{N-N_{\rm t}}{S}+\frac{N_{\rm t}}{S_0} < \rho < \frac{N}{S}.
\label{eq:rhos}
\eeq

\begin{figure}
\centering
\includegraphics[width=\columnwidth]{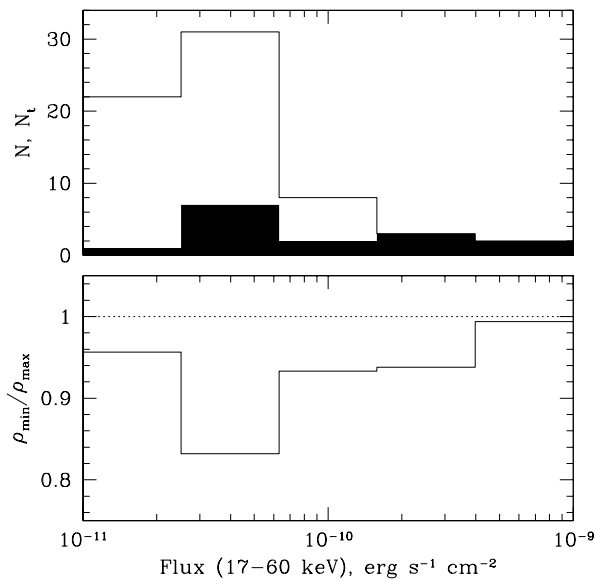}
\caption{{\sl Top:} Differential distribution of hard X-ray fluxes for
all INTEGRAL emission-line AGN at $|b|>5^\circ$ (upper histogram) and
only for those targeted by INTEGRAL (filled histogram). {\sl
Bottom:} The ratio of the minimum and maximum estimates of the sky
density of sources according to Eq.~(\ref{eq:rhos}).
}
\label{fig:target_effect}
\end{figure}

Our calculation of the luminosity function was implicitly based on
using $\rho_{\rm max}$. Figure~\ref{fig:target_effect}
shows the ratio $\rho_{\rm min}/\rho_{\rm max}$ as a function of flux
for our AGN survey. It can be seen that the effect under consideration
is strongest at medium fluxes
($\sim$~2--6~$10^{-11}$~erg~s$^{-1}$~cm$^{-2}$) where the fraction of
targeted AGN is relatively high, while the sky coverage is
relatively low. However, even at the maximum the effect is less than 17\%. This
value can be considered an upper limit on the possible overestimation of the
luminosity function due to the presence of targeted AGN in the
sample. We note that implementing observations of ``empty''
extragalactic regions with INTEGRAL (Proposal 0320108) has
allowed us to significantly reduce the non-randomness of the survey.  

We may conclude from the above that the total systematic uncertainty
in the luminosity function due to source identification incompleteness
and AGN pointings is probably $\lesssim 10$\% (taking into account that
the two effects tend to counterbalance each other), which is less than
the present statistical uncertainties.

\begin{figure}
\centering
\includegraphics[width=\columnwidth]{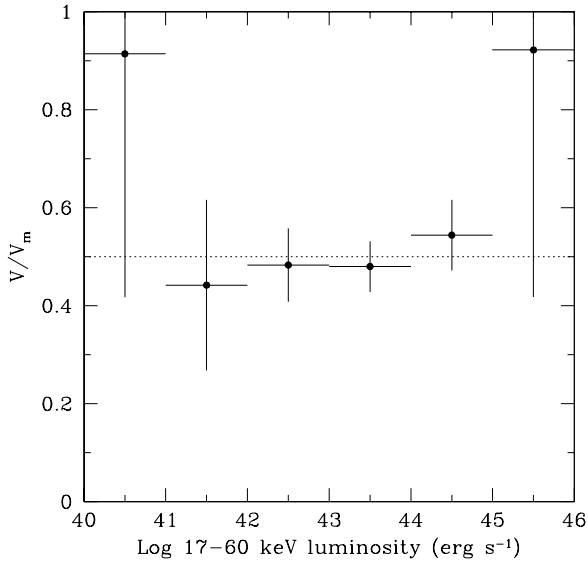}
\caption{$V/\vm$ ratio averaged over luminosity bins for the AGN
sample used for the construction of the luminosity function. The error
bars represent 1$\sigma$ statistical uncertainties.
}
\label{fig:lum_vvmax_bin}
\end{figure}

Finally, the standard $V/\vm$ test \citep{schmidt68} applied to our
AGN sample demonstrates (Fig.~\ref{fig:lum_vvmax_bin}) that the
spatial distribution of the studied AGN population is consistent with
being homogeneous.

\subsection{Comparison with the 3--20~keV luminosity function}
\label{s:3-20}

It is interesting to compare the local AGN luminosity function 
obtained with INTEGRAL in the hard X-ray band (17--60~keV)
with that previously constructed in the adjacent, medium X-ray band
(3--20~keV) using the RXTE Slew Survey \citep{sazrev04}. 

\begin{figure}
\centering
\includegraphics[bb=60 160 440 730, width=\columnwidth]{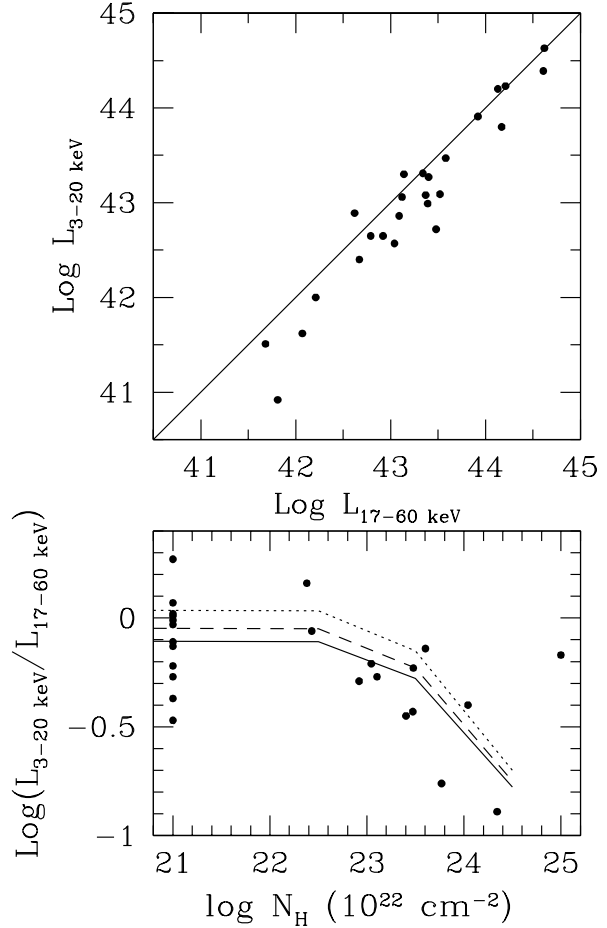}
\caption{{\sl Top:} The 3--20~keV luminosity as a function of
17--60~keV luminosity for AGN detected during both the RXTE Slew
Survey and INTEGRAL all-sky survey. The line indicates equality of the
two luminosities. {\sl Bottom:} The 3--20~keV/17--60~keV luminosity
ratio as a function of $\nh$ (points). The
data shown at $\log\nh=21$ represent unobscured  
AGN ($\log\nh<22$). Also shown are expected luminosity ratios in the
absence of source variability for an intrinsic (unabsorbed) spectrum
consisting of a power law ($\Gamma=1.8$) and a Compton reflection
component with relative amplitude $R=0$ (dotted line), $R=0.5$ (dashed
line), and $R=1$ (solid line).
}
\label{fig:integral_xss}
\end{figure}

For this comparison we need to know how the 17--60~keV
luminosities ($\lhx$) of Seyfert galaxies are related to their 3--20~keV
luminosities ($\lx$). One possibility is to directly compare $\lhx$
with $\lx$ for those emission-line AGN detected during both the XSS
and the INTEGRAL all-sky survey. We have in total 25 such cases, for
which we show in Fig.~\ref{fig:integral_xss} $\lx$ vs. $\lhx$ and the 
ratio $\lx/\lhx$ as a function of $\nh$. These data are compared with
the expected dependence of $\lx/\lhx$ on $\nh$ assuming that
the intrinsic (unabsorbed) AGN spectrum consists of a power law with a
photon index $\Gamma=1.8$ plus a Compton reflection component (XSPEC
model pexrav, \citealt{magzdz95}) of different relative amplitudes $R$
($\equiv\Omega/2\pi$, for a zero inclination angle). As expected (see
the upper panel of Fig.~\ref{fig:integral_xss}), for most of the AGN
$\lx<\lhx$ mainly due to the presence of intrinsic X-ray absorption
and (to a lesser degree) of Compton reflection. The measured $\lx/\lhx$
ratios are apparently consistent with the expected $\nh$ trend, 
somewhat more so in the case of a non-negligible reflection component
($R\sim$~0.5--1). The observed significant (a
factor of $\sim 2$) scatter of the data around this trend can be naturally
explained by intrinsic variability of the sources, since the RXTE and
INTEGRAL observations are separated by years.

Taking the outcome of this test into account and considering the abundant
literature on the broad-band X-ray spectra of Seyfert galaxies 
(e.g. \citealt{gonetal96,peretal02}), we may adopt to a first
approximation that local emission-line AGN have a universal intrinsic 
spectrum consisting of a power law with $\Gamma=1.8$ and a
substantial reflection component ($R\sim$~0.5--1). We can then
estimate the average ratio $\langle\lx/\lhx\rangle$ for low-
($\log\lhx<43.6$) and high- ($\log\lhx>43.6$) luminosity AGN taking
their (different) 
observed $\nh$ distributions (Fig.~\ref{fig:nhdist_lum}) into account. For the
low-luminosity AGN, we find $\langle\lx/\lhx\rangle=0.79$ (0.67) for
$R=0.5$ (1). For the high-luminosity AGN, most of which are
unobscured, the corresponding values are 0.96 (0.84). Note that, in
this context, changing the reflection amplitude $R$ by 0.5
corresponds to changing the power-law slope by $\sim 0.1$. Given the
above numbers, we may adopt to a first approximation that 
$\log\langle\lhx/\lx\rangle\approx 0.1$ regardless of luminosity, and
accordingly convert the XSS luminosity function to the
17--60~keV band.
 
\begin{figure}
\centering
\includegraphics[width=\columnwidth]{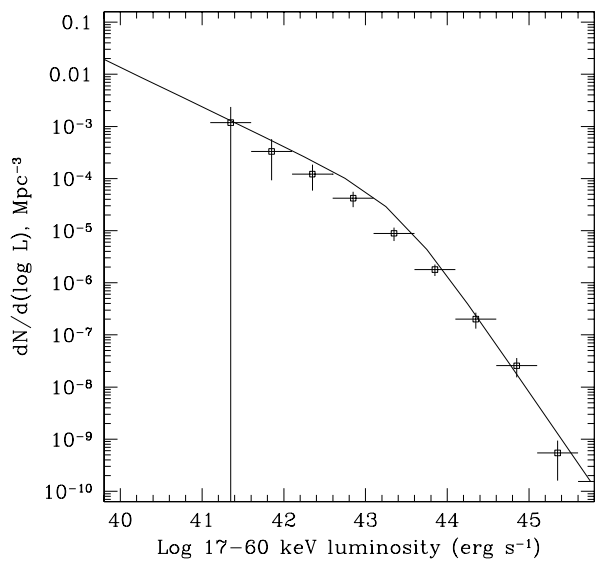}
\caption{The best-fit model of the hard X-ray AGN luminosity
function obtained with INTEGRAL (line) in comparison with the
luminosity function based on the RXTE Slew Survey in the
3--20~keV band \citep{sazrev04}, here converted to the 17--60~keV band
(points with error bars).
} 
\label{fig:lumfunc_xss}
\end{figure}

The result is shown in Fig.~\ref{fig:lumfunc_xss} in comparison with the
hard X-ray luminosity function measured directly by INTEGRAL. Note
that the amplitude and error bars for the XSS luminosity function take
into account the systematic uncertainty due to possible incompleteness
of the XSS sample of AGN \citep{sazrev04}. The two luminosity
functions agree with each other, as is formally
confirmed by a comparison of the parameters of the best-fit analytic
models. However, there is an indication that the XSS luminosity function lies 
somewhat below the INTEGRAL one. Here it is best to
consider the cumulative luminosity density. With RXTE we measured
$\rho_{\rm 3-20~keV}(>41)=(5.2\pm 1.2)\times
10^{38}$~erg~s$^{-1}$~Mpc$^{-3}$, which for the
$\langle\lhx/\lx\rangle$ ratio adopted above implies that
$\rho_{\rm 17-60~keV}(>41.1)=(6.5\pm 1.5)\times
10^{38}$~erg~s$^{-1}$~Mpc$^{-3}$. With INTEGRAL we found  
$\rho_{\rm 17-60~keV}(>41)=(12.4\pm 1.5)\times
10^{38}$~erg~s$^{-1}$~Mpc$^{-3}$, i.e. a factor of $1.9\pm 0.5$ higher
luminosity density.

\section{Implications for the cosmic X-ray background}
\label{s:cxb}

It has become a common paradigm that the bulk of the cosmic X-ray
background (CXB) is composed of emission from all AGN in the
Universe. This conclusion is mainly based on the fact that deep
extragalactic X-ray surveys have resolved $\sim 80$\% of the CXB at
energies below $\sim 8$~keV \citep{hicmar06}. However, the maximum of
the CXB $\nu F_\nu$ spectrum is located at $\sim 30$~keV, and at these
energies not more than $\sim 3$\% of the CXB has been resolved into
point sources in the deepest exposures with INTEGRAL (and much less by previous
missions). Therefore, resolving the CXB near its peak will be one of
the main tasks of future hard X-ray missions. For the time being, it
is interesting to address the following question: is the X-ray
absorption distribution measured with INTEGRAL for the local AGN population
consistent with that required for distant quasars to explain the CXB
spectrum? 

To answer this question, we can deduce relative fractions of AGN with
different absorption columns from the measured local $\nh$
distribution (Fig.~\ref{fig:nhdist_lum}): 
\beqa
\log\nh&<22.0:&~ 0.37\pm 0.10,
\nonumber\\
22.0<\log\nh&<22.5: &~ 0.20\pm 0.07,
\nonumber\\
22.5<\log\nh&<23.0: &~ 0.05\pm 0.04,
\nonumber\\
23.0<\log\nh&<23.5: &~ 0.13\pm 0.06,
\nonumber\\
23.5<\log\nh&<24.0: &~ 0.16\pm 0.07,
\nonumber\\
24.0<\log\nh&<24.5: &~ 0.07\pm 0.05,
\nonumber\\
\log\nh&>24.5: &~ 0.02\pm 0.02.
\label{eq:weights}
\eeqa
These fractions take into account that low- ($\log\lhx<43.6$)
and high- ($\log\lhx>43.6)$ luminosity AGN contribute $\sim 90$\% and
$\sim 10$\%, respectively, to the local luminosity density. The quoted
errors result from Poissonian statistics and we have ignored the systematic
uncertainty associated with the lack of $\nh$ measurements for
several AGN in our sample.

We next adopt a fiducial intrinsic AGN spectrum consisting of a
cutoff power-law component and the corresponding Compton
reflection component with relative amplitude $R$:
\beq
\frac{dN_\gamma}{dE}=AE^{-\Gamma}\exp(-E/E_{\rm f})+R f(E),
\label{eq:fidspec}
\eeq
where $f(E)$ is again described by the pexrav model \citep{magzdz95}.
We adopt $\Gamma=1.8$, $E_{\rm f}=200$~keV, and $R=0.5$ or $R=1$ as
reference values.  

\begin{figure}
\centering
\includegraphics[width=\columnwidth]{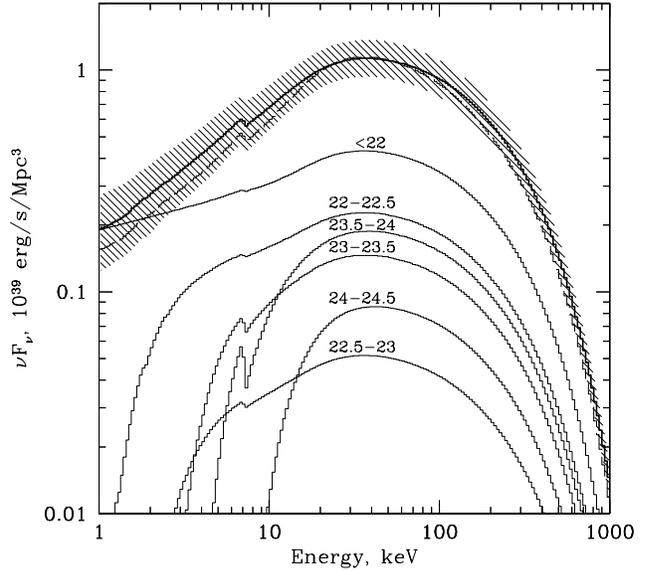}
\caption{Composite spectrum of the local emission-line AGN obtained
assuming that the intrinsic AGN spectrum is given by
Eq.~(\ref{eq:fidspec}) with $\Gamma=1.8$, $E_{\rm f}=200$~keV, 
and $R=0.5$ (thick solid line), and taking the observed $\nh$
distribution of nearby AGN (Fig.~\ref{fig:nhdist_lum})into
account. The spectrum is normalized to the measured local AGN
17--60~keV luminosity density. The shaded area indicates the
uncertainty in the composite 
spectrum due to the uncertainties in the $\rho_{\rm 17-60~keV}(>40)$
value and in the $\nh$ distribution. Also shown are the contributions
to the composite spectrum of AGN with different degrees of obscuration
(curves with labels indicating $\log\nh$). The dashed line shows the
composite spectrum in the case of $R=1$.
}
\label{fig:composite_spectrum}
\end{figure}

We can now build the composite spectrum of the local AGN by
propagating the intrinsic spectrum given by Eq.~(\ref{eq:fidspec})
through different absorption columns and  
summing up the resulting spectra with the weights introduced 
in Eq.~(\ref{eq:weights}). The resulting spectrum is shown in
Fig.~\ref{fig:composite_spectrum}. Note that we have neglected the
contribution of substantially Compton-thick AGN ($\log\nh>24.5$) since
it is expected to be small ($\sim 3$\%) given the observed $\nh$
distribution (Fig.~\ref{fig:nhdist_lum}), whereas the spectra of
Compton-thick AGN can assume quite different shapes depending on the
geometry of the obscuring and reflecting material (e.g. \citealt{matetal00}). 
 
The composite spectrum shown in Fig.~\ref{fig:composite_spectrum} has
been normalized so as to reproduce the local AGN luminosity
density measured with INTEGRAL, $\rho_{\rm 17-60~keV}(>40)$
(Table~\ref{tab:lumfunc}), i.e. this spectrum represents the volume
emissivity of all local emission-line AGN with $\log\lhx>40$. In
Fig.~\ref{fig:composite_spectrum} we indicate the range of uncertainty in the
composite spectrum resulting from the uncertainties in the $\nh$
distribution and in the $\rho_{\rm 17-60~keV}(>40)$ value. It can also be seen
that the difference between the normalized composite spectra for $R=0.5$ and
$R=1$ is only noticeable below $\sim 15$~keV.

Suppose now that the same spectral shape characterizes AGN at all
redshifts and that only the AGN luminosity density evolves with
$z$. Then the cumulative AGN emission observed at $z=0$ will have a spectrum
(for a flat cosmology, e.g. \citealt{sazetal04})
\beq
I(E)=\frac{c}{4\pi H_0}\int_0^\infty\frac{\epsilon(z) F((1+z)E)}
{(1+z)\left[\Omega_{\rm m}(1+z)^3+\Omega_\Lambda\right]^{1/2}}dz,
\label{eq:bgr}
\eeq
where $F(E)$ is the template spectrum shown in
Fig.~\ref{fig:composite_spectrum} and $\epsilon(z)$ is a function
describing the evolution of the luminosity density.

\begin{figure}
\centering
\includegraphics[width=\columnwidth]{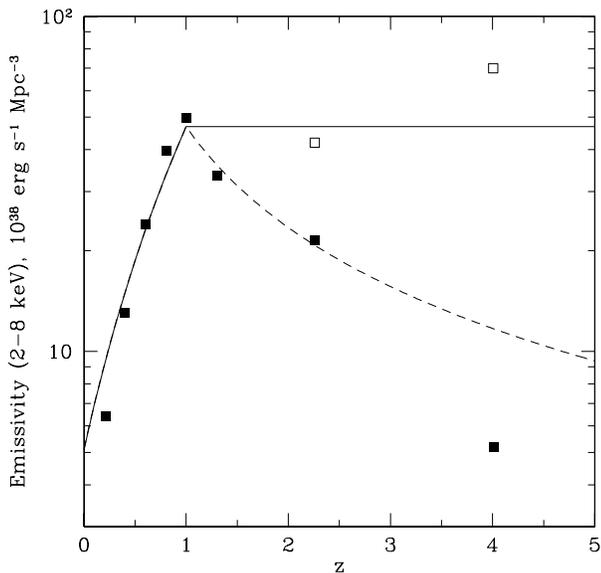}
\caption{Evolution with redshift of the rest-frame 2--8~keV luminosity
density based on Chandra extragalactic surveys (adopted from
\citealt{baretal05}). The measurements shown by the filled squares
take only spectroscopically identified AGN into account, while those
shown by the empty squares also take the maximal
incompleteness into account. We assume that the true evolution 
is bound between the two limiting functions given by
Eqs.~(\ref{eq:e1}) and (\ref{eq:e2}), shown by the dashed and
solid lines, respectively.
}
\label{fig:evol_models}
\end{figure}

The best constraints on $\epsilon(z)$ have so far been coming from
deep X-ray surveys with Chandra and XMM-Newton. In
Fig.~\ref{fig:evol_models} we reproduce the results of
\cite{baretal05} on the redshift dependence of the rest-frame 2--8~keV
luminosity density of AGN, which reflects the observed pure luminosity
evolution of the AGN luminosity function at $z\lesssim
1.5$. Between $z\sim 0.2$ and $z\sim 1$ the evolution is
well-constrained and is consistent with  
a power-law dependence ($z^\alpha$) with a slope $\alpha\sim
3.2$. At $z\gtrsim 1$ there is still significant uncertainty due to
the incompleteness of source identification and redshift
determination, and the evolution function $\epsilon(z)$ is bound by
the two sets of data points in Fig.~\ref{fig:evol_models}, or
approximately by the following two limiting functions, also shown in
the figure:
\beqa
e_1(z)\propto 
\left\{
\begin{array}{ll}
(1+z)^{3.2},\,\,& z\le 1\\
e_1(1)/z,\,\,& z>1
\end{array}
\right.
\label{eq:e1}
\eeqa
\beqa
e_2(z)\propto
\left\{
\begin{array}{ll}
(1+z)^{3.2},\,\,& z\le 1\\
e_2(1),\,\,& z>1.
\end{array}
\right.
\label{eq:e2}
\eeqa

\begin{figure}
\centering
\includegraphics[width=\columnwidth]{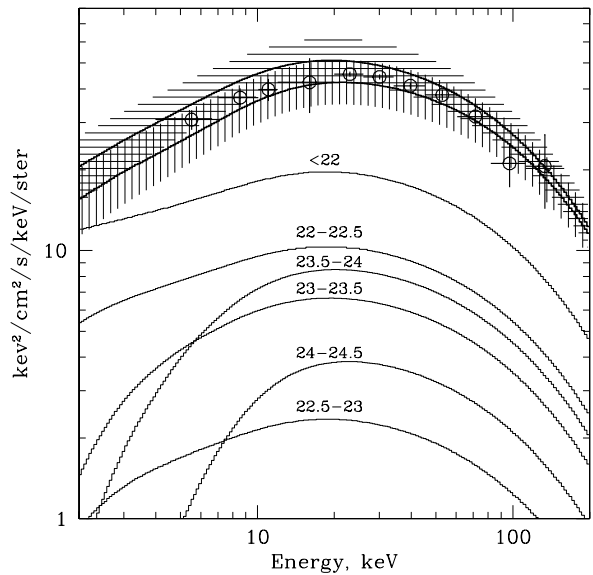}
\caption{Predicted CXB spectrum based on the local AGN composite spectrum
(Fig.~\ref{fig:composite_spectrum}, for $R=0.5$) and the AGN redshift
evolution measured with Chandra (Fig.~\ref{fig:evol_models}). The
upper thick solid line corresponds to the scenario of flat AGN
luminosity density evolution at $z>1$, the horizontally shaded region
indicating the corresponding uncertainty. The lower solid line and the
vertically shaded region correspond to the scenario of $\propto 1/z$
evolution at $z>1$. The other presented curves show the contributions
of AGN with different degrees of obscuration (the labels indicate
$\log\nh$). The points with error bars show the CXB spectrum measured
with the JET-X, IBIS/ISGRI, and SPI instruments on INTEGRAL
\citep{chuetal06}.
\label{fig:cxb_spectrum}
}
\end{figure}

By convolving our locally-determined composite AGN spectrum
(Fig.~\ref{fig:composite_spectrum}) with these two alternative
evolution laws, we obtain the cumulative AGN spectra shown in
Fig.~\ref{fig:cxb_spectrum}. The shaded regions around these spectra
represent the superposition of the present uncertainties in the local
$\rho_{\rm 17-60~keV}(>40)$ value and in the local $\nh$ distribution. 
For comparison in Fig.~\ref{fig:cxb_spectrum} the CXB
broad-band spectrum is shown as recently measured with the different
instruments on INTEGRAL \citep{chuetal06}. It can be seen that the
predicted cumulative AGN spectrum agrees very well, both in
amplitude and in shape, with the CXB for both limiting scenarios of AGN
evolution at $z>1$. It should be noted, though, that our calculation was
based on the INTEGRAL estimate of the cumulative luminosity density
of local AGN with $\log\lhx>40$, so the inclusion of even lower
luminosity AGN and normal galaxies could somewhat increase
the normalization of the predicted cumulative AGN spectrum.  

We conclude that the combination of data from the INTEGRAL all-sky
survey and deep extragalatic X-ray surveys seems to be able to
naturally explain the observed CXB spectrum. Furthermore these data are
consistent with the cosmological evolution of AGN at $z\lesssim 1.5$
(where the bulk of the CXB is produced) having occurred mostly in
luminosity and much less (if any) in spectral properties such as the
X-ray absorption distribution (for a given value of $\lhx/\lb(z)$).

\section{Conclusions}
\label{s:summary}

We have used the INTEGRAL all-sky hard X-ray survey to
study some key properties of the local ($z\lesssim 0.1$) AGN
population. Since our source detection was based on 17--60~keV fluxes,
the survey is equally sensitive to AGN with X-ray absorption
columns up to several $10^{24}$~cm$^{-2}$, i.e. well into
the Compton-thick domain ($\nh>1.5\times 10^{24}$~cm$^{-2}$). 

One of the most surprising results of the survey is that very
few Compton-thick AGN have been detected, and all of them were  
known before. The observed fraction of Compton-thick
objects is only $\sim 10$\%. This estimate may increase to at most $\sim
20$\% once the currently missing absorption columns are measured. Clearly
Compton-thick AGN are fairly rare. A note of caution is necessary here
concerning very Compton-thick objects ($\log\nh\sim$~25--26). Since
their observed hard X-ray luminosities are expected to be $\lesssim 10$\% of
their intrinsic luminosities, such sources may constitute a major
fraction of AGN at luminosities near our survey's effective 
limit ($\lhx\sim 10^{40}$--$10^{41}$~erg~s$^{-1}$) and below.

Another major result of our study is confirmation of the result of
the RXTE Slew Survey that the local AGN absorption distribution is very
different at low ($\log\lhx<43.6$) and high ($\log\lhx>43.6$)
luminosities. While the fraction of obscured ($\log\nh>22$) objects
is $\sim 70$\% among the low-luminosity AGN, it is only $\sim 25$\%
among the high-luminosity ones. We note that a similar result was
recently reported based on a combination of different HEAO-1 (2--10~keV)
surveys \citep{shietal06} and is also emerging from the all-sky hard X-ray
survey carried out by Swift  \citep{maretal05}. Furthermore, a similar
trend has been found at higher redshifts (e.g. \citealt{uedetal03}). 

We measured the hard X-ray luminosity distribution of local
emission-line AGN. Its broken power-law shape is in good agreement with recent 
determinations in softer energy bands, in particular at 3--20~keV
\citep{sazrev04} and at 2--10~keV \citep{shietal06}. We also found the
cumulative luminosity density of AGN with $\lhx>10^{41}$~erg~s$^{-1}$
to be $\rho_{\rm 17-60~keV}(>41)=(12.4\pm 1.5)\times
10^{38}$~erg~s$^{-1}$~Mpc$^{-3}$. This is a factor of $\sim 1.9$
higher than the XSS measurement in the 3--20~keV band assuming that
typically $\langle L_{\rm 17-60~keV}/L_{\rm 3-20~keV}\rangle\sim
1.25$. Thus the two determinations differ by $\sim 2.8\sigma$,
although there is a significant uncertainty in the average ratio
$\langle L_{\rm 17-60~keV}/L_{\rm 3-20~keV}\rangle$. Our 
INTEGRAL estimate of the AGN cumulative luminosity 
density in the 17--60~keV band also seems to be consistent with the 2--10~keV
estimate of \cite{shietal06}: $\rho_{\rm 2-10~keV}(>42)=(7.2\pm
1.4)\times 10^{38}$~erg~s$^{-1}$~Mpc$^{-1}$ (for 
$H_0=75$~km~s$^{-1}$~Mpc$^{-1}$), given that on average $L_{\rm 
17-60~keV}/L_{\rm 2-10~keV}\sim 2$. Finally, the hard X-ray
luminosity function obtained in this work agrees with the one constructed
by \citet{becetal06} based on a smaller sample of INTEGRAL AGN,
improving it in terms of accuracy and reducing the bias associated
with AGN pointed observations. 

In summary, the all-sky X-ray and hard X-ray surveys performed with
RXTE, INTEGRAL, Swift, and HEAO-1 have provided an accurate census
of nearby unobscured and obscured AGN. This new information is not
only interesting in its own right but also provides a reliable $z=0$
point for studies of AGN cosmological evolution and the growth of
massive black holes. 

As a first attempt to apply the local AGN statistics in a cosmological
context, we demonstrated that the spectral shape and amplitude of
the CXB allow for the possibility that the absorption
distribution of AGN (for a given $\lhx/\lb(z)$ ratio) has not changed
significantly since $z\sim 1.5$, while the AGN luminosity function 
has experienced pure luminosity evolution. It should be noted that we came
to this tentative conclusion using a fiducial (although realistic)
intrinsic AGN spectrum. In future work we plan to perform a more
self-consistent analysis by using INTEGRAL data on the hard X-ray
spectra of nearby AGN. We also note that it is possible that the hard
X-ray spectra of nearby Seyferts are in reality somewhat different 
from those of distant powerful quasars, since to a first approximation
the intrinsic (unabsorbed) spectrum is expected to depend on the black
hole mass, accretion rate \citep{shasun76}, and spin. Observations
with future hard X-ray telescopes will permit direct tests of whether
the hard X-ray spectra of quasars are similar to those of local
Seyferts or not.

\smallskip
\noindent {\sl Acknowledgments} This work was supported by the
DFG-Schwerpunktprogramme (SPP 1177). The research made 
use of the NASA/IPAC Extragalactic Database (operated by the Jet
Propulsion Laboratory, California Institute of Technology), SIMBAD
database (operated at the CDS, Strasbourg), and data of different X-ray
astronomy missions obtained through the High Energy Astrophysics
Science Archive Research Center Online Service, provided by the
NASA/Goddard Space Flight Center. Some of the INTEGRAL sources were
identified thanks to follow-up X-ray observations with Chandra and 
Swift/XRT. INTEGRAL is an ESA project funded by ESA member 
states (especially the PI countries: Denmark, France, Germany, Italy,
Spain, Switzerland), Czech Republic and Poland, and with the
participation of Russia and the USA.


\end{document}